\documentclass[12pt]{article}

 \usepackage{amssymb}
\usepackage{amsfonts}
 \usepackage{amsmath}
\usepackage{graphicx}
\usepackage{cite}

\newcommand{\be}{\begin{equation}}
\newcommand{\ee}{\end{equation}}
\newcommand{\bea}{\begin{eqnarray}}
\newcommand{\eea}{\end{eqnarray}}

\newcommand{\lb}{\label}
\newcommand{\p}[1]{(\ref{#1})}

\newcounter{rown}

\begin{document}

\begin{titlepage}

\vspace*{1.5cm}

\renewcommand{\thefootnote}{\dag}
\begin{center}

{\LARGE\bf Supersymmetrization of the 3-particle }

\vspace{0.45cm}

{\LARGE\bf elliptic Calogero model}

\vspace{2.5cm}

{\large\bf Sergey Fedoruk}
 \vspace{1.5cm}

{\it Bogoliubov Laboratory of Theoretical Physics, }\\
{\it Joint Institute for Nuclear Research,}\\
{\it 141980 Dubna, Moscow region, Russia} \\
\vspace{0.1cm}

{\tt fedoruk@theor.jinr.ru}

\vspace{1.5cm}

\end{center}

\vspace{1.0cm} \vskip 0.6truecm \nopagebreak

\begin{abstract}
\noindent
$\mathcal{N}{=}\,2$ and $\mathcal{N}{=}\,4$ supersymmetric generalizations of the 3-particle elliptic Calogero system are proposed.
Supersymmetry generators of the system are found in which the center of mass sector is described by the supermultiplet
${\bf (1,\mathcal{N},\mathcal{N}-1)}$, while the sector of relative coordinates is the supermultiplet ${\bf (2,\mathcal{N},\mathcal{N}- 2)}$.
The $\mathcal{N}{=}\,2$ model with three supermultiplets ${\bf (1,2,1)}$ is also presented.
\qquad

\end{abstract}

\vspace{3cm}

\newpage

\end{titlepage}

\setcounter{footnote}{0}
\setcounter{equation}0
\section{Introduction}

The important role of integrable many-particle Calogero systems \cite{C,OP,Poly-rev,Ar} in the study of various models in various areas of theoretical physics suggests the natural problem of constructing supersymmetric generalizations of such systems.
For the rational, trigonometric and hyperbolic Calogero systems, various types of supersymmetrizations were found, both for different types of root systems and for a different number of supersymmetries (see, for example, \cite{FrMen,Wyl,BGK,GLK,FIL08,superc,FIL19,KLPS19 ,Fed20,KLS19,KL20,Fed20a} and references there).
With the elliptic Calogero system, the situation is fundamentally different: there are still no supersymmetric elliptic Calogero systems for $n\geq3$ particles.\footnote{The 2-particle Calogero system reduces to separated one-dimensional subsystems of the center of mass and relative motion, and supersymmetrization of each of them does not cause difficulties.}

The previously constructed supersymmetric generalizations of $n$-particle Calogero systems with $\mathcal{N}$ supersymmetries used $n$ so-called ${\bf (1,\mathcal{N},\mathcal{N}-1)}$-supermultiplets, each containing one physical bosonic degree of freedom and $\mathcal{N}$ physical fermionic degrees of freedom, while the $\mathcal{N}-1$ bosonic degrees of freedom of the corresponding supermultiplet were auxiliary. That is, under $\mathcal{N}$-supersymmetrization of the $n$-particle Calogero system, the bosonic phase space with coordinates $q_a$, $a=1,\ldots,n$ and momenta $p_a$ was expanded by $n\mathcal{ N}$ real Grassmann variables $\psi_a^A=(\psi_a^A)^*$, $A=1,\ldots,\mathcal{N}$. Such a procedure in the case of  the elliptic Calogero system with $n\geq3$ particles encounters difficulties: the required properties, which are satisfied in the rational, trigonometric and hyperbolic cases for functions in supercharges, are not satisfied for the chosen functions in the elliptic case.

In this short paper, we propose another way of constructing supersymmetric generalizations of the elliptic Calogero model using the example of a 3-particle system with coordinates $q_1,q_2,q_3$. The proposed supersymmetrization procedure is not limited by the requirement to use ${\bf (1,\mathcal{N},\mathcal{N}-1)}$-supermultiplets for a supersymmetric description of each of the coordinates. Moreover, supersymmetrization of the center of mass sector with the coordinate $(q_1+q_2+q_3)/\sqrt{3}$ and the sector whose configuration space is parameterized by the variables $q_a-q_b$,\footnote{For brevity, the $q_a-q_b$ variables will be called below relative or difference coordinates, and the dynamic sector described by such coordinates will be called the relative sector of the system.} is performed independently and is described by different supermultiplets in the general case. This procedure, described for $\mathcal{N}{=}\,2$ and $\mathcal{N}{=}\,4$ supersymmetrizations in Sections 2 and 3, respectively, is carried out in terms of new bosonic variables that are functions of the original coordinates $q_1,q_2,q_3$ and their momenta. In addition to the fact that the resulting supercharges have a complicated dependence on the coordinates of the bosonic phase space, a distinctive feature of the proposed supersymmetrization is the use of the ${\bf (2,2,0)}$ supermultiplet in one of the possible $\mathcal{N}{=}\, 2$ systems and the supermultiplet ${\bf (2,4,2)}$ in the $\mathcal{N}{=}\,4$ case to describe the sector of relative coordinates.

\setcounter{equation}0
\section{3-particle elliptic Calogero model}

The Hamiltonian of the 3-particle elliptic Calogero system in the case of the root system $A_{2}$ has the form \cite{C,OP,Ar}
\be \lb{L-3b-1}
H_{3b}\ =\ \frac12\ \Big[(p_1)^2 + (p_2)^2 + (p_3)^2 \Big] \ +\  (2g)^2 \Big[ \wp(q_{12})+\wp(q_{23})+\wp(q_{31})\Big] \,  ,
\ee
where $q_{a}$, $p_{a}$, $a=1,2 ,3$ form canonical pairs of the phase space: their non-zero Poisson brackets are $\{q_{a}, p_{b} \}=\delta_{ab}$. The variables $q_{ab}:=q_{a}-q_{b}$ are relative (difference) coordinates of the configuration space. In \p{L-3b-1}, the function $\wp(z)=\wp(z|\omega,\omega^\prime)$ is the second-order Weierstrass elliptic function with periods $2\omega$ and $2\omega^ \prime$ \cite{BeEr}, while the constant $g$ characterizes the interaction of particles (for the compactness of the following expressions, the interaction constant in the Hamiltonian \p{L-3b-1} is given as a factor $(2g)^2$).

Using the addition theorem for the $\wp$-function \cite{BeEr}
\be \lb{sum-P}
\wp(z_1)+\wp(z_2)+\wp(z_1+z_2)\ =\ \frac14 \left[\frac{\wp^{\prime}(z_1)-\wp^{\prime}(z_2)}{\wp(z_1)-\wp(z_2)}\right]^2 \,  ,
\ee
its parity $\wp(-z)=\wp(z)$ and condition on the relative coordinates $q_{23}+q_{31}=-q_{12}$, we represent the Hamiltonian \p{L-3b-1} as
\be \lb{L-3b-2}
H_{3b}\ =\ \frac12\, \Big[(p_1)^2 + (p_2)^2 + (p_3)^2 \Big] \ +\
g^2 \left[\frac{\wp^{\prime}(q_{13})+\wp^{\prime}(q_{23})}{\wp(q_{13})-\wp(q_{23})}\right]^2 \,  ,
\ee

Let us move on to new coordinates, for which we take the coordinate of the center of mass 
\be \lb{coord-cm}
x_0 = \big(q_{1}+q_{2}+q_{3}\big)/\sqrt{3}
\ee
and the Jacobi coordinates (see, for example, \cite{C})
\be \lb{coord-Ja}
x_1 = q_{12}/\sqrt{2}\,,\qquad x_2 = \big(q_{13}+q_{23}\big)/\sqrt{6}\,  .
\ee
Expressions \p{coord-Ja} imply
\be \lb{coord-Ja-inv}
q_{12} = \sqrt{2}\, x_1\,,\qquad q_{13}= \big(\sqrt{3}\,x_2+x_1\big)/\sqrt{2} \,,\qquad
q_{23}= \big(\sqrt{3}\,x_2-x_1\big)/\sqrt{2}\,  .
\ee
New momenta $\pi_0$, $\pi_1$, $\pi_2$ that have non-zero Poisson brackets $\{x_{0}, \pi_0 \}=\{x_1, \pi_1 \}=\{x_2, \pi_2 \}=1$ are defined by the expressions
\be \lb{mom-Ja}
\pi_0 = \big(p_{1}+p_{2}+p_{3}\big)/\sqrt{3}\,,\qquad \pi_1= p_{12}/\sqrt{2} \,,\qquad
q_{23}= \big(p_{13}+p_{23}\big)/\sqrt{6}\,  ,
\ee
where $p_{ab}:=p_{a}-p_{b}$.
In the variables \p{coord-cm}, \p{coord-Ja}, \p{mom-Ja} the Hamiltonian \p{L-3b-2} takes the form
\be \lb{L-3b-3}
H_{3b}\ =\ \frac12\,(\pi_0)^2 \ +\ \frac12\,\Big[(\pi_1)^2 + (\pi_2)^2 \Big] \ +\
g^2 \left[
\frac{\wp^{\prime}\Big(\frac{\displaystyle\sqrt{3}x_2+x_1}{\displaystyle\sqrt{2}}\Big)+
\wp^{\prime}\Big(\frac{\displaystyle\sqrt{3}x_2-x_1}{\displaystyle\sqrt{2}}\Big)}
{\wp\Big(\frac{\displaystyle\sqrt{3}x_2+x_1}{\displaystyle\sqrt{2}}\Big)-
\wp\Big(\frac{\displaystyle\sqrt{3}x_2-x_1}{\displaystyle\sqrt{2}}\Big)}\right]^2 \,  .
\ee
The last term in \p{L-3b-3} is represented as the derivative of the logarithm. As a result of this, we have
\be \lb{L-3b-4}
H_{3b}\ =\ \frac12\,(\pi_0)^2 \ +\ \frac12\,\Big[(\pi_1)^2 + (\pi_2)^2 \Big] \ +\
2g^2 \left[
{\frac{\partial}{\partial x_1}}
\ln\left\{\textstyle\wp\Big(\frac{\displaystyle\sqrt{3}x_2+x_1}{\displaystyle\sqrt{2}}\Big)-
\wp\Big(\frac{\displaystyle\sqrt{3}x_2-x_1}{\displaystyle\sqrt{2}}\Big)\right\}
\right]^2 \,  .
\ee

In the system described by the Hamiltonian \p{L-3b-4}, the center of mass sector with phase coordinates $x_0$, $\pi_0$ has split off and describes free motion in this direction. The supersymmetrization of this sector is trivial. Thus, for complete supersymmetrization of the system, it is necessary to supersymmetrize the two-particle sector with coordinates $x_1$, $x_2$ and momenta $\pi_1$, $\pi_2$.

\setcounter{equation}{0}
\section{$\mathcal{N}{=}\,2$ supersymmetrization}

To obtain $\mathcal{N}{=}\,2$ supersymmetrization of the 3-particle system under consideration, we pass in the sector with coordinates $x_1,x_2$ (or dependent relative coordinates $q_a-q_b$) to the complex phase variables
\be \lb{coord-compl}
z= \big(x_1+ix_2\big)/\sqrt{2} \,,\quad \bar z= \big(x_1-ix_2\big)/\sqrt{2} \,,\qquad
p_z= \big(\pi_1-i\pi_2\big)/\sqrt{2} \,,\quad \bar p_z= \big(\pi_1+i\pi_2\big)/\sqrt{2} \,,
\ee
the nonzero Poisson brackets of which are equal to $\{z,p_z\}=\{\bar z,\bar p_z\}=1$.

In terms of the new variables \p{coord-compl}, the Hamiltonian \p{L-3b-3} becomes
\be \lb{H-3b-3}
H_{3b}\ =\ \frac12\,(\pi_0)^2 \ +\ p_z\bar p_z \ +\
g^2 \big[V(z,\bar z)\big]^2 \,  ,
\ee
where
\be \lb{V-def}
V(z,\bar z)\ :=\
\frac{\wp^{\prime}(\displaystyle e^{-i\pi/3}z+e^{i\pi/3}\bar z)-
\wp^{\prime}(\displaystyle e^{i\pi/3}z+e^{-i\pi/3}\bar z)}
{\wp(\displaystyle e^{-i\pi/3}z+e^{i\pi/3}\bar z)-
\wp(\displaystyle e^{i\pi/3}z+e^{-i\pi/3}\bar z)} \,  .
\ee
Note that the following expression takes place:
\be \lb{V-W}
V(z,\bar z)\ =\
\left(\frac{\partial}{\partial z}+\frac{\partial}{\partial \bar z}\right) \ln|W(z,\bar z)|\,  ,
\ee
where
\be \lb{W-def}
W(z,\bar z)\ :=\
{\wp(\displaystyle e^{-i\pi/3}z+e^{i\pi/3}\bar z)-
\wp(\displaystyle e^{i\pi/3}z+e^{-i\pi/3}\bar z)} \,  .
\ee

The next step before the symmetrization procedure is
the introduction of polar coordinates in the momentum space parametrized by the variables $p_z$ and $\bar p_z$. That is, instead of $p_z(t)$ and $\bar p_z(t)$, we introduce the variables $r(t)$ and $\alpha(t)$, defined by the relations
\be \lb{pz-rho-def}
p_z=r e^{i\alpha} \,,\qquad \bar p_z=r e^{-i\alpha} \,  .
\ee
Then the expressions for the coordinates $z(t)$ and $\bar z(t)$ in the new variables have the form:
\be \lb{z-rho-def}
z=- e^{-i\alpha} (2r)^{-1}\left(r p_r -i p_\alpha\right)\,,\qquad
\bar z=- e^{i\alpha} (2r)^{-1}\left(r p_r +i p_\alpha\right) \,  ,
\ee
where $p_r$ and $p_\alpha$ are the momentum variables for $r$ and $\alpha$: the non-zero Poisson brackets of the variables $r$, $\alpha$, $p_r$, $p_\alpha$ are
\be \lb{PB-r}
\{r,p_r\}=\{\alpha,p_\alpha\}=1 \,  .
\ee
Since, in terms of the new phase coordinates, the variables used in \p{W-def} are
\be \lb{comb-z}
\begin{array}{rcl}
e^{-i\pi/3}z+e^{i\pi/3}\bar z&=&\displaystyle - \cos\!\big(\alpha+\frac{\pi}{3}\big)p_r
+r^{-1}\sin\!\big(\alpha+\frac{\pi}{3}\big)p_\alpha\,,\\ [6pt]
e^{i\pi/3}z+e^{-i\pi/3}\bar z&=&\displaystyle - \cos\!\big(\alpha-\frac{\pi}{3}\big)p_r
+r^{-1}\sin\!\big(\alpha-\frac{\pi}{3}\big)p_\alpha \,  ,
\end{array}
\ee
the function $V=V(r,\alpha,p_r,p_\alpha )=V(z(r,\alpha,p_r,p_\alpha ),\bar z(r,\alpha,p_r,p_\alpha )) $ obtained by substituting variables \p{z-rho-def} into \p{V-def} becomes
\be \lb{V-n}
V =
-\frac{\wp^{\prime}\Big(\displaystyle \cos\!\big(\alpha+\frac{\pi}{3}\big)p_r-r^{-1}\sin\!\big(\alpha+\frac{\pi}{3}\big)p_\alpha\Big)-
\wp^{\prime}\Big(\displaystyle \cos\!\big(\alpha-\frac{\pi}{3}\big)p_r-r^{-1}\sin\!\big(\alpha-\frac{\pi}{3}\big)p_\alpha\Big)}
{\wp\Big(\displaystyle \cos(\alpha+\frac{\pi}{3})p_r-r^{-1}\sin(\alpha+\frac{\pi}{3})p_\alpha\Big)-
\wp\Big(\displaystyle \cos(\alpha-\frac{\pi}{3})p_r-r^{-1}\sin(\alpha-\frac{\pi}{3})p_\alpha\Big)} \,  .
\ee
As a result, the Hamiltonian of the 3-particle Calogero \p{H-3b-3} system in the new variables becomes
\be \lb{H-3b-3a}
H_{3b}\ =\ \frac12\,(\pi_0)^2 \ +\ r^2 \ +\
g^2 \big[V(r,\alpha,p_r,p_\alpha)\big]^2 \,  .
\ee
In a sense, the canonical change of variables $(z,\bar z; p_z, \bar p_z)\to (r,\alpha; p_r, p_\alpha)$ in the relative coordinate sector, defined in \p{pz-rho-def}, \p{z-rho-def}, can be interpreted as a transition to a ``dual'' equivalent system with the coordinate equal to the radial part of the complex momentum.

Let us show that the system with the Hamiltonian \p{H-3b-3a} can be embedded into the $\mathcal{N}{=}\,2$ supersymmetric system as a bosonic subsystem of the latter.

\subsection{$\mathcal{N}{=}\,2$ system using ${\bf (2,2,0)}$ multiplet}

Consider first the case of $\mathcal{N}{=}\,2$ supersymmetry with a minimum number of Grassmann variables used.

In addition to the bosonic phase variables $x_0$, $\pi_0$, $r$, $\varphi$, $p_r$, $p_\varphi$, we introduce a pair of complex Grassmann variables $\chi$, $\bar\chi= (\chi)^*$ and $\psi$, $\bar\psi=(\psi)^*$, whose nonzero (graded) Poisson brackets (or Dirac brackets in Lagrangian theory) are equal to
\be \lb{PB-gr}
\{\chi,\bar\chi\}=-i \,,\qquad
\{\psi,\bar\psi\}=-i \,.
\ee
The variable $\chi$ serves for supersymmetrization of the center of mass sector, then the use of $\psi$ will allow constructing a supersymmetric generalization of the sector of relative coordinates.

The supercharges 
\bea \lb{Q-2}
Q&=& \pi_0\chi \ + \ \sqrt{2}\,\Big[r \ + \ ig\,V(r,\alpha,p_r,p_\alpha)\Big]\psi\,, \\ [6pt]
\bar Q&=& \pi_0\bar\chi \ + \ \sqrt{2}\,\Big[r  \ - \ i  g\, V(r,\alpha,p_r,p_\alpha)\Big]\bar\psi
\lb{bQ-2}
\eea
form, with respect to the graded Poisson bracket \p{PB-r}, \p{PB-gr}, the $\mathcal{N}{=}\,2$ supersymmetry algebra:
\be \lb{super2}
\{Q,\bar Q\}=-2i H\,,\qquad
\{Q,Q\}=\{\bar Q,\bar Q\}\,,
\ee
where the supersymmetric Hamiltonian 
\be \lb{H-sup-3}
H \ =\ H_{3b} \ + \ 2g\,\psi\bar\psi \,\frac{\partial}{\partial p_r}\,V(r,\alpha,p_r,p_\alpha)
\ee
has the part $H_{3b}$ that is independent of the Grassmann variables and coincides with the Hamiltonian \p{H-3b-3} of the 3-particle elliptic Calogero system.
Note that due to relations \p{super2} and the Jacobi identities, the Hamiltonian \p{H-sup-3} has zero Poisson brackets with supercharges \p{Q-2}, \p{bQ-2}:
\be \lb{super2a}
\{H,Q\}=\{H,\bar Q\}=0\,.
\ee
Thus, the system described by the Hamiltonian \p{H-sup-3} and the supercharges \p{Q-2}, \p{bQ-2} is a $\mathcal{N}{=}\,2$ supersymmetrization of the system with the Hamiltonian \p{H-3b-3} describing the 3-particle elliptic Calogero system.

The structure of supercharges \p{Q-2}, \p{bQ-2} shows that the center of mass sector with bosonic phase variables $(x_0, \pi_0)$ is described by the $\mathcal{N}{=}\,2$ supermultiplet ${\bf (1,2,1)}$, while the relative sector of the remaining bosonic variables $(r, \varphi; p_r, p_\varphi)$ (or equivalently $(x_1,x_2; \pi_1,\pi_2) $) in a supersymmetric system is given by the $\mathcal{N}{=}\,2$ supermultiplet ${\bf (2,2,0)}$. The last statement about the $\mathcal{N}{=}\,2$ supermultiplet ${\bf (2,2,0)}$ is based on the number of bosonic and fermionic physical degrees of freedom in the relative sector and their variation under transformations generated by supercharges \p{Q-2}, \p{bQ-2}.

Note that the fermionic sector of the model with the Hamiltonian \p{H-sup-3} and supercharges \p{Q-2}, \p{bQ-2} is 4-dimensional, in contrast to the $\mathcal{N}{=}\,2$ supersymmetric Calogero models in the non-elliptic cases \cite{FrMen,BGK,FIL08,superc,FIL19,KLPS19,Fed20,KLS19} using at least 6 fermions in the 3-particle case.

\subsection{$\mathcal{N}{=}\,2$ model in terms of ${\bf (1,2,1)}$ multiplets}

Another variant of $\mathcal{N}{=}\,2$ supersymmetrization takes place when a pair of fermion variables is introduced for all bosonic degrees of freedom.
That is, in the supersymmetrization of the sector of relative coordinates, twice as many fermion variables are used as in the previous case.
That is, in addition to the fermions $\chi$, $\bar\chi=(\chi)^*$ and $\psi$, $\bar\psi=(\psi)^*$ of the previous subsection, we introduce one more pair of complex Grassmann variables $\varphi$, $\bar\varphi=(\varphi)^*$, whose non-zero (graded) Poisson brackets are
\be \lb{PB-gra}
\{\varphi,\bar\varphi\}=-i \,.
\ee
For a system extended by three complex Grassmann variables, one can construct two types of supercharges commuting on the Hamiltonian whose bosonic part is equal to \p{H-3b-3}.

The first type is the supercharges
\bea \lb{Q-2-4}
Q&=& \pi_0\chi \ + \ (r  +  ig\,V)\psi \ + \ (r  -  ig\,V)\varphi  \ - \
\frac{g}{r}\frac{\partial V}{\partial p_r}\big(\varphi\bar\varphi\psi-\psi\bar\psi\varphi\big)\,, \\ [6pt]
\bar Q&=& \pi_0\bar\chi \ + \ (r  -  ig\,V)\bar\psi \ + \ (r  +  ig\,V)\bar\varphi  \ - \
\frac{g}{r}\frac{\partial V}{\partial p_r}\big(\varphi\bar\varphi\bar\psi-\psi\bar\psi\bar\varphi\big)\,,
\lb{bQ-2-4}
\eea
which, with respect to the Poisson brackets, form the superalgebra \p{super2}, \p{super2a}, where the supersymmetric Hamiltonian has the form
\be \lb{H-sup-4}
H \ =\ H_{3b} \ + \ 2g\,\frac{\partial V}{\partial p_r}\,\big(\psi\bar\psi- \varphi\bar\varphi\big) \ + \
2g^2\Big\{V,\frac{1}{r} \frac{\partial V}{\partial p_r}\Big\}\psi\bar\psi\varphi\bar\varphi\,.
\ee

The second type of generators $\mathcal{N}{=}\,2$ of the superalgebra \p{super2}, \p{super2a} is the supercharges 
\bea \lb{Q-2-5}
Q&=& \pi_0\chi \ + \ (r  +  ig\,V)\psi \ + \ (r  -  ig\,V)\varphi  \ - \
i\frac{\partial \ln V}{\partial p_r}\big(\varphi\bar\varphi\psi+\psi\bar\psi\varphi\big)\,, \\ [6pt]
\bar Q&=& \pi_0\bar\chi \ + \ (r  -  ig\,V)\bar\psi \ + \ (r  +  ig\,V)\bar\varphi  \ + \
i\frac{\partial \ln V}{\partial p_r}\big(\varphi\bar\varphi\bar\psi+\psi\bar\psi\bar\varphi\big)
\lb{bQ-2-5}
\eea
and the Hamiltonian
\be \lb{H-sup-5}
H \ =\ H_{3b} \ + \ 2g\,\frac{\partial V}{\partial p_r}\,\big(\psi\bar\psi+ \varphi\bar\varphi\big) \ - \
2\,\frac{\partial^2 \ln V}{\partial p_r^{\,2}}\,\psi\bar\psi\varphi\bar\varphi\,.
\ee

Unlike supercharges \p{Q-2}, \p{bQ-2}, supercharges \p{Q-2-4}, \p{bQ-2-4} and \p{Q-2-5} , \p{bQ-2-5} contain terms of the third degree in fermions.
As a result of this, the Hamiltonian \p{H-sup-5} has the term of the fourth degree in fermions, unlike the Hamiltonian \p{H-sup-3}.

\setcounter{equation}{0}
\section{3-particle $\mathcal{N}{=}\,4$ elliptic Calogero model}

By analogy with the $\mathcal{N}{=}\,2$ case considered in Subsection 3.1, one can construct the charges of the $\mathcal{N}{=}\,4$, $d\,{=}\,1 $ Poincar\'{e} superalgebra with the Hamiltonian whose bosonic part coincides with the Hamiltonian of the 3-particle elliptic Calogero system \p{H-3b-3}.

In addition to the variables $\mathcal{N}{=}\,2$ in the case of Subsection 3.1, we introduce the second pairs of Grassmann variables.
That is, we consider a system with fermions $\chi_i$, $\bar\chi^i=(\chi_i)^*$ and $\psi_i$, $\bar\psi^i=(\psi_i)^*$, $i=1,2$, whose nonzero Poisson brackets are
\be \lb{PB-gra4}
\{\chi_i,\bar\chi^j\}=-i \delta_i^j\,,\qquad\{\psi_i,\bar\psi^j\}=-i \delta_i^j\,.
\ee

Within the framework of the procedure considered in the previous section, two possible variants of $\mathcal{N}{=}\,4$ supercharges are found.

In the first case, the generators $\mathcal{N}{=}\,4$, $d\,{=}\,1$ of the Poincare superalgebra have the form:
\bea \lb{Q-4a}
Q_i&=& \pi_0\chi_i \ + \ \sqrt{2}\left[(r  +  ig\,V)\psi_i  \ - \
\frac{g}{r}\frac{\partial V}{\partial p_r}\,\bar\psi^k\psi_k\psi_i\right], \\ [6pt]
\bar Q^i&=& \pi_0\bar\chi^i \ + \ \sqrt{2}\left[(r  -  ig\,V)\bar\psi^i \ - \
\frac{g}{r}\frac{\partial V}{\partial p_r}\,\bar\psi^k\psi_k\bar\psi^i\right],
\lb{bQ-4a}
\\ [6pt]
\lb{H-4a}
H&=& H_{3b} \ - \ 2g\,\frac{\partial V}{\partial p_r}\,\bar\psi^k\psi_k \ - \
g^2\Big\{V,\frac{1}{r} \frac{\partial V}{\partial p_r}\Big\}\,(\bar\psi^k\psi_k)^2\,,
\eea
where $H_{3b}$ is the Hamiltonian of the 3-particle elliptic Calogero system given in \p{H-3b-3}.
It is easy to check that quantities \p{Q-4a}, \p{bQ-4a}, \p{H-4a} have the Poisson brackets 
\be \lb{super4}
\{Q_i,\bar Q^j\}=-2i H\,\delta_i^j\,,\qquad
\{Q_i,Q_j\}=\{\bar Q^i,\bar Q^j\}
\ee
and, as a consequence of \p{super4} and the Jacobi identities, the Poisson brackets
\be \lb{super4a}
\{H,Q_i\}=\{H,\bar Q^i\}=0
\ee
hold.
 
Another variant $\mathcal{N}{=}\,4$, $d\,{=}\,1$ of the Poincar\'{e} superalgebra \p{super4}, \p{super4a} is realized by the generators
\bea \lb{Q-4b}
Q_i&=& \pi_0\chi_i \ + \ \sqrt{2}\left[(r  +  ig\,V)\psi_i  \ - \
i\frac{\partial \ln V}{\partial p_r}\,\bar\psi^k\psi_k\psi_i\right], \\ [6pt]
\bar Q^i&=& \pi_0\bar\chi^i \ + \ \sqrt{2}\left[(r  -  ig\,V)\bar\psi^i \ + \
i\frac{\partial \ln V}{\partial p_r}\,\bar\psi^k\psi_k\bar\psi^i\right],
\lb{bQ-4b}
\\ [6pt]
\lb{H-4b}
H&=& H_{3b} \ - \ 2g\,\frac{\partial V}{\partial p_r}\,\bar\psi^k\psi_k \ + \
\frac{\partial^2 \ln V}{\partial p_r^{\,2}}\,(\bar\psi^k\psi_k)^2\,.
\eea

Introduced in \p{Q-4a}, \p{bQ-4a}, \p{H-4a} and \p{Q-4b}, \p{bQ-4b}, \p{H-4b} the $\mathcal{N}{=}\,4$ systems are described in the center of mass sector by the coordinate $x_0$ and 4 fermions $\chi_i$, $\bar\chi^i$.
That is, this sector is described by the ${\bf (1,4,3)}$ multiplet.
At the same time, the description of the relative sector with two coordinates $r$, $\alpha$ also uses four fermions $\psi_i$, $\bar\psi^i$.
This shows that this sector is given by the $\mathcal{N}{=}\,4$ multiplet ${\bf (2,4,2)}$.
Similar to the $\mathcal{N}{=}\,2$ case of Section 3.1, the number of fermions in the $\mathcal{N}{=}\,4$ systems \p{Q-4a}, \p{bQ-4a} , \p{H-4a} and \p{Q-4b}, \p{bQ-4b}, \p{H-4b} equal to 8 is less than the number of fermions used (at least 12 in the 3-particle case) in the $\mathcal{N}{=}\,4$ systems of the papers \cite{Wyl,FIL08,superc,FIL19,KLPS19,Fed20,KL20}.

\setcounter{equation}{0}
\section{Final remarks}

In this paper, $\mathcal{N}{=}\,2$ and $\mathcal{N}{=}\,4$ supersymmetric generalizations of the 3-particle elliptic Calogero system are presented.
An important point in the construction was going beyond the use of the ${\bf (1,2,1)}$ and ${\bf (1,4,3)}$ multiplets in the $\mathcal{N}{=}\,2 $ and $\mathcal{N}{=}\,4$ cases, respectively.
Moreover, $\mathcal{N}{=}\,4$ supersymmetrization uses both the ${\bf (1,4,3)}$ and ${\bf (2,4,2)}$ multiplets.
In this case, supersymmetrization of the relative sector was actually performed in bosonic coordinates, which are functions of the initial coordinates $q_a$ and momenta $p_a$.

This paper does not present expressions for supercharges in terms of the initinal variables.
This is easily obtained by inverse transformations of the coordinates in the phase space, but how to use the obtained cumbersome expressions is still unclear.
In this regard, no superfield description of the resulting supersymmetric systems has been found.

Of course, the ``non-synchronous in all initial coordinates $q_a$'' supersymmetrization used in the paper can also be applied to 3-particle Calogero systems of other types: rational, trigonometric, and hyperbolic.
The situation with the generalization of the application of the supersymmetrization procedure for all types of Calogero systems is similar to the $\mathcal{N}{=}\,2$ supersymmetric model considered in \cite{KLS19}.\footnote{The application of the system obtained in \cite{KLS19} to describe the $\mathcal{N}{=}\,2$ elliptic Calogero system was not discussed, although this is possible within the framework of the approach considered there.}
In \cite{KLS19}, the $n$-particle Hamiltonian had a structure similar to that considered in this paper, but its $\mathcal{N}{=}\,2$ supersymmetrization required an increased number ($2n^2$) of fermions, in contrast with the models considered here with their reduced number.
This paper is one of the attempts to find the most appropriate supersymmetrization procedure for the Calogero model in the elliptic case.
The answer to this will be given by considering supersymmetrizations of elliptic Calogero systems with different numbers of supersymmetries and with a number of particles greater than three, which we plan to study in the future.

\smallskip
\section*{Acknowledgments}
I would like to thank Evgeny Ivanov and Sergey Krivonos for useful discussions and comments.
This work was supported by the Russian Science Foundation grant no. 21-12-00129.

\end{document}